# SYNERGISTIC NANOPHOTONICS OF FULLERENE

E.F.Sheka

Peoples' Friendship University of Russia, 117197 Moscow, Russia

sheka@icp.ac.ru

**Abstract**: A deep similarity of photo-stimulated effects occurring in physical and biological objects involving fullerene forces to raise the question: what is meant under nanophotonics of fullerene and if should we not imply under this conventional term, usually restricted to optical events, something more general? Discussed in the paper, makes it possible to suggest that the formation of positive-negative fullerene ion pair at each photon absorption act is common for photo stimulated events in chemistry, medicine, and optics thus providing their common origin.

## 1. INTRODUCTION

The photo activity of fullerenes is mainly manifested itself via three groups of photosensitive phenomena, among which there are
- Photostimulated chemical reactions including a particular reaction of dimerization and/or oligomerization of fullerenes
- Photodynamic effect in the therapy of diluted fullerene aqueous solutions
- Photoinduced enhancement of spectral properties of fullerene solutions.

In spite of that these phenomena have different appearance and are related to different scientific topics, there is a strong feeling of a common origin of the fullerene behavior concerning all of them. Discussed in the paper, makes it possible to suggest that the formation of positive-negative fullerene ion pair at each photon absorption act is common for all the events and evidently provides their common origin. In view of this, the three phenomena differ only by the manner of this action implementation. The creation of the ionic pair is a consequence of peculiarities in the intermolecular interaction (IMI) between fullerene molecules aggravated by a significant contribution of the donor-acceptor component.

## 2. Context on Intermolecular Interaction in $C_{60}$-Based Binary Systems

The first issue related to the problem concerns the necessity to examine the configuration interaction of the states of neutral molecules and their ions for the IMI terms of both ground and excited states of the system to be constructed [1]. Generally, the IMI term of a D-A system, $E_{int}(r, R)$, is a sum of two terms, namely, $E_{int}(A^+B^-)$ and $E_{int}(A^0B^0)$, and is complicated function of intra- and intermolecular coordinates.

The term $E_{int}(A^+B^-)$ describes the interaction of ions leading to their bonding in point $R^{+-}$. Similarly, the term $E_{int}(A^0B^0)$ bounds neutral molecules in point $R^{00}$.

The second states that at short intermolecular distances $E_{int}(A^+B^-)$ is always below $E_{int}(A^0B^0)$. However, at longer distances the situation may change. The term $E_{int}(A^0B^0)$ tends on infinity to its asymptotic limit $E_{inf}(A^0B^0)$ equal to zero. The asymptotic limit of the term $E_{int}(A^+B^-)$ is equal to $I_A - \varepsilon_B$, where $I_A$ determines ionization potential (IP) of molecule $A$, and $\varepsilon_B$ is the electron affinity (EA) of molecule $B$. If $I_A - \varepsilon_B > 0$, terms $E_{int}(A^+B^-)$ and $E_{int}(A^0B^0)$ intersect.

The third one deals with the configuration interaction between the states of neutral molecules and their ions that provides avoiding the intersection so that in the vicinity of $R_{scn}$ the terms split forming two branches of combined IMI terms, lower branches of which describe the ground state of the system while the upper ones are related to excited states.

A two-well shape of the IMI term of the ground state is the main feature of a D-A system pointing to a possible existence of two stable structural configurations formed by the system constituents. The configurations involve weakly bound molecular complexes $A+B$ at comparatively large intermolecular distances $R^{00}$ and new chemical products $AB$ at intermolecular distances $R^{+-}$ compared to the lengths of chemical bonds. It should be noted that actually each IMI term represents a surface in the multi-dimensional space, therefore the existence of several minima, both in the region $R^{+-}$ and $R^{00}$, cannot be excluded.

A practical realization of both $A+B$ and $AB$ products depends on a particular shape of the IMI term, two of four possible types [2] of which are shown in Fig. 1. The IMI term of type 1 (Fig. 1a) implies that both products are energetically stable. The $[A^+B^-]$ adduct ($AB$) formed at $R^{+-}$ has a clear ionic origin, so that its structure and electron properties are determined mostly by interaction of molecular ions. In contrast to this, the neutral molecules of the system are responsible for the properties of the $[A^0B^0]$ complex ($A+B$) in the vicinity of $R^{00}$.

A particular photosensitivity of the $C_{60}$-based binary system is connected with phototransitions related to the $A+B$ complexes that in the dominant majority of cases are charge transfer complexes [2]. If the light excitation length $\lambda_{exc}$ lies within the $B_2$ absorption band shown by arrow in Fig.1a, the photoexcitation results in the transferring the complex $A+B$ of neutral molecules to a pair of molecular ions $A^+ + B^-$, whose further behavior determines which kind of photosensitive phenomena will be observed. Let us consider three scenarios that are influenced by such photoexcitation, namely

- Transformation of the $A+B$ complex into the relevant $AB$ adduct
- Changing interaction of the $A+B$ complex with other molecular species under photoexcitation
- Enhancement of the $A+B$ complex optical response.

## 3. Photosensitive Chemical Reactions

Evidently the first of the above scenarios concerns D-A chemical reactions. Among the latter two limiting classes are connected with the two IMI terms shown in Fig.1.

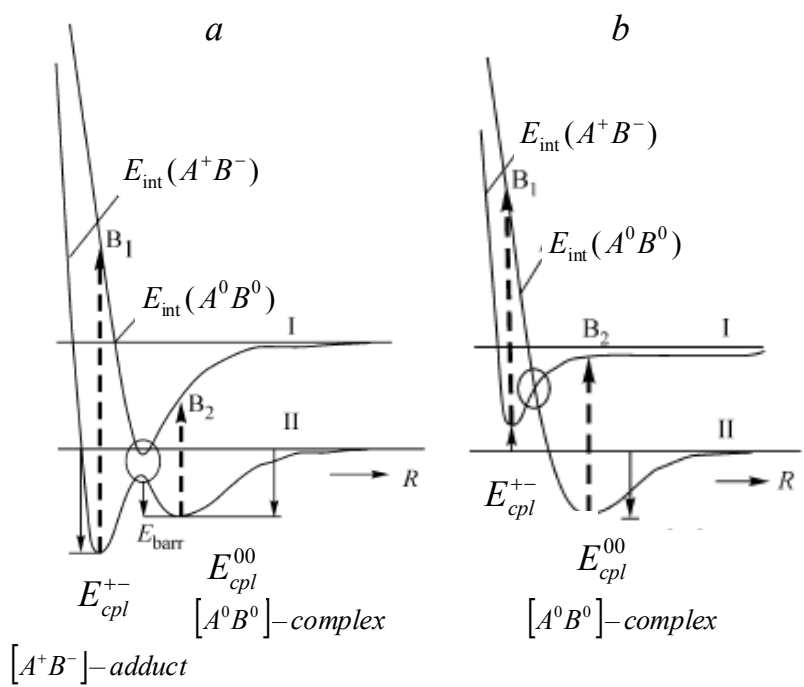

**Figure 1.** Scheme of intermolecular interaction terms of type 1 and 3 [2]. *a.* $\left|E_{cpl}^{+-}\right| > \left|E_{cpl}^{00}\right|$; *b.* $E_{cpl}^{+-} > 0$ and $E_{cpl}^{00} < 0$. Ovals match regions of the term intersection; $B_1$ and $B_2$ mark phototransitions from the corresponding minima of the ground state terms.

In the case of IMI term of type 1(Fig.1*a*), both coupling energies are negative and $\left|E_{cpl}^{+-}\right| > \left|E_{cpl}^{00}\right|$ so that the minimum at $R^{+-}$ dominates. Obviously, D-A interaction plays a governing role leading to the formation of *AB* adduct. Evidently, the reaction starts with the formation of the *A+B* complex. Passing to *AB* one is possible when overcoming a barrier. There are many empirical ways how to overcome the barrier. The topic is considered in details in Ref. 3 particularly exemplified by the dimerization of fullerene $C_{60}$. Here is worthwhile to mention only a photoexcitation that willingly promotes the transformation from the *A+B* stage to *AB* one. As said previously, excitation in the spectral region covering $B_2$ absorption band causes ionization of the neutral molecules. Coulomb interaction between the ions afterwards facilitates passing through the intersection region to the minimum at $R^{+-}$ [4]. Photosensitized reactions of this kind are well known (see [5] and references therein).

Among the latter photodimerization (as well as photooligomerization) of both $C_{60}$ [3, 6] and $C_{70}$ [6, 7] fullerenes should be mentioned particularly. Figure 2 presents a concise report on the photostimulated dimerization and/or oligomerization of fullerene $C_{60}$ as it follows from [3]. A cluster consisting of five molecules, which is formed at intermolecular spacing of ~3-5Å corresponding to $R^{00}$ minima in Fig.1*a* [8], acquires one positive-negative ion pair at the first act of the light absorption. The energy of this ion-pair-incorporating cluster is much above the barrier of the $C_{60}$ molecule dimerization [3] and the Coulomb interaction between the ions readily provides their chemical bonding via 2, 2-cycloaddition that is characteristic for $C_{60}$ dimer. As shown [3], dimer $(C_{60})_2$ preserves donor-acceptor characteristics typical for individual $C_{60}$ molecules, so that its interaction with the monomer molecule is subordinated to the same interaction term in Fig.1a and the next act of the light absorption will produce the creation of the ion pair of either $[(C_{60})_2^- - C_{60}^+]$ or $[(C_{60})_2^+ - C_{60}^-]$ composition. The Coulomb interaction promotes the formation of trimer $(C_{60})_3$ incorporating the latter in the remaining cluster. Since the trimer possesses the same donor-acceptor properties as monomer and previously dimer, the photoexcitation of cluster with such trimer will produce the formation of one of possible ion pairs such as $[(C_{60})_3^- - C_{60}^+]$, $[(C_{60})_3^- - (C_{60})_2^+]$ as well as those with inverted polarities of ions so that the oligomerization proceeds. The monomer donor-acceptor characteristics typical for $C_{60}$ oligomer of arbitrary composition allows for presenting the photostimulated oligomerization in practice as a series of $[(C_{60})_n^- - (C_{60})_m^+]$ [2, 2]-cycloaddition bondings governed by a pairwise interaction similar to that between monomer molecules.

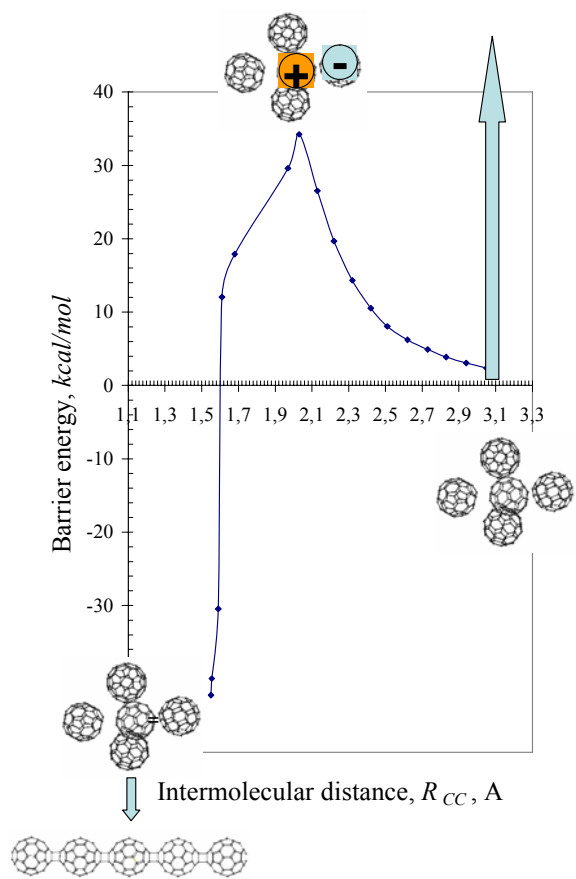

**Figure 2**. A sketch of photostimulated oligomerization of fullerene $C_{60}$. The barrier energy is taken from [3]. Atomic compositions of $(C_{60})_5$ cluster and oligomer correspond to equilibrium structures of the relevant AM1 UBS HF solutions [8, 9].

In contrast to homoconstituent photostimulated reactions considered above, the heteroconstituent ones can be best of all exemplified by interaction of $C_{60}$ with amines [2, 10]. Figure *3a* shows the equilibrium structures of the *A+B* and *AB* products formed in due course of interaction of $C_{60}$ with pyridine-dimethylenemethylamine (*PDMMA*). The constituents interaction is subordinated to the IMI term shown in Fig.1a [2] and two adducts are formed at different initial spacing between the components. The complex $A+B$ is typical charge transfer one while the $AB$ adduct shows all properties related to the $C_{60}$ amine monoderivatives [10]. The photoexcitation of solutions containing these and similar constituents effectively facilitates the formation of the corresponding $AB$ adducts in practice [5].

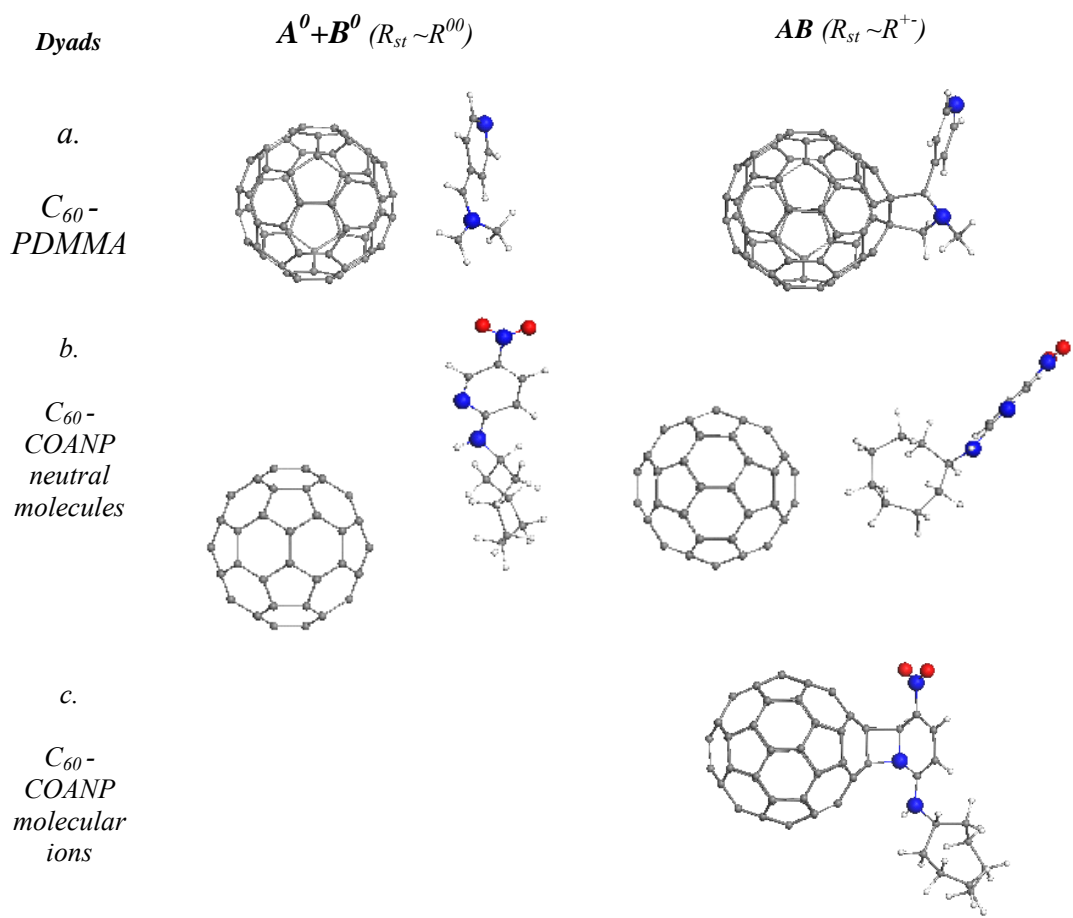

**Figure 3**. Equilibrium structures of $C_{60}$-based dyads with two amines, namely, pyridine-dimethylenemethylaniline (PDMMA) [2] and 2-cyclooctylamine-5-nitropyridine (COANP) [10] at two starting distances. $R_{st}$ marks the shortest initial intermolecular spacing between the dyad component before the structure optimization. UBS HF AM1 singlet state. Carbon atoms are shown by small dark balls. Big blue and red balls mark nitrogen and oxygen atoms. Hydrogen atoms are shown by small white balls.

However, IMI term of type 1 is characteristic not for all amines. The relationship between $E_{cpl}^{+-}$ and $E_{cpl}^{00}$ depends on the amine structure so that the case when $E_{cpl}^{+-} > 0$ while $E_{cpl}^{00} < 0$ cannot be excluded. Changing the coupling energy sign indicates weakening the D-A interaction and decreasing its contribution to the total IMI term. At the same time when the inequality $E_{cpl}^{+-} > \left|E_{cpl}^{00}\right|$ is not too strong, the IMI term of the ground state can still be two-well but with a definite preference to the $R^{00}$ minimum against the $R^{+-}$ one. Under these conditions the formation of $AB$ product from the $A+B$ complex is energetically non-profitable. However, if the molecules are preliminary ionized from the $A+B$ state, the formed ions may form a stable product $AB$. Figures 3*b* and 3*c* demonstrate equilibrium structures of the binary system

$C_{60}$+COANP [11] showing that a chemically bound composition is formed if only both components are ionized. Reactions of this kind are often called as hidden photochemical reactions [4]. This is the very reaction that causes changing color of fullerene solutions when they are stored without precautions against the light illumination. Obviously, the bigger the inequality $E_{cpl}^{+-} > |E_{cpl}^{00}|$, the less probable is the formation of the *AB* product from an energetically stable *A*+*B* complex. When additionally to the inequality the $R^{+-}$ minimum is rather shallow or is absent at all, the IMI term takes one-well shape constructed predominantly of term $E_{\text{int}}(A^0B^0)$. This is the limiting case when the D-A interaction contribution is the weakest.

## 4. Photostimulated Spin-Flip and the Photodynamic Therapeutic Effect of Fullerene Solutions

The most interesting phenomenon concerning changing interaction of fullerene with other molecular species under photoexcitation according to scenario 2 is known as the photodynamic therapeutic effect of fullerene solution [12]. It concerns the oxidative action of fullerene solutions in both molecular and polar solvents in the presence of molecular oxygen. As accepted, the action consists in the oxidation of targets by singlet oxygen $^1O_2$ that is produced in due course of photoexcitation of fullerene solutions involving convenient triplet oxygen $^3O_2$.

The difference in the behavior of singlet and triplet oxygen is obviously connected with the difference in the pairing of the molecule electrons caused by different spin multiplicity. A quantitative characteristic of the pairing can be expressed in terms of the total number of unpaired electrons $N_D$ that can be calculated by using unrestricted broken symmetry Hartree-Fock approximation [13]. Calculations performed within the framework of the AM1 version of semiempirical UBS HF approach expose $N_D$=2 for both spin states. But, if for the triplet state this finding just naturally reflects two electrons that are responsible for maintaining the molecule spin multiplicity, the availability of two effectively unpaired electrons in the singlet state evidences a biradical character of the molecule, which explains $^1O_2$ high oxidative activity. Therefore, the photostimulated $^3O_2 \rightarrow {}^1O_2$ transformation in the presence of fullerene molecule just means exempting the molecule two electrons from the spin multiplicity service thus transforming chemically inactive molecule into a biradical.

The presence of fullerene for the photostimulated $^3O_2 \rightarrow {}^1O_2$ transformation is ultimately necessary so that the treatment was called as photodynamic fullerene therapy [14, 15]. For the reason alone that the action is provided by a complex involving fullerene and solvent molecules as well as molecular oxygen, it becomes clear that it is resulted from a particular intermolecular interaction. However, until now, the mechanism of the photodynamic effect has been hidden behind a slogan 'triplet state photochemical mechanism' that implies the excitation transfer over a chain of molecules according to a widely accepted scheme [16-18]

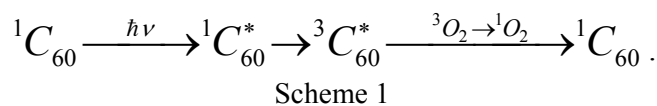

Scheme 1

The scheme implies the energy transfer from the singlet photoexcited fullerene to the triplet one that further transfers the energy to convenient triplet oxygen thus transforming the latter into singlet oxygen. The first two stages of this 'single-fullerene-molecule' mechanism are quite evident while the third one, the most important for the final output, is quite obscure in spite of a lot of speculations available [17-19]. Obviously, the stage efficacy depends on the strength of the intermolecular interaction between fullerene and oxygen molecules. Numerous quantum chemical calculations show that pairwise interaction in the dyad $[C_{60}+O_2]$ in both singlet and triplet state is practically absent. The AM1 UBS HF computations performed in the current paper fully support the previous data disclosing the coupling energy of the dyad $E_{cpl}^{f-o}$ equal to zero in both cases. This puts a serious problem for the explanation of the third stage of the above scheme forcing to suggest the origination of a peculiar intermolecular interaction between $C_{60}$ and $O_2$ molecules in the excited state once absent in the ground state.

However, the intermolecular interaction in the photodynamic (PD) solutions is not limited by the fullerene-oxygen (*f-o*) interaction only. There are two other interactions, namely: fullerene-fullerene (*f-f*) and fullerene-solvent (*f-s*), among which the former is quite significant thus revealing itself in the fullerene dimerization [3]. The *f-s* interaction in the case of aqueous and benzene solutions can be so far ignored. Besides a significant strength, the *f-f* interaction possesses peculiar features caused by the exclusive D-A ability of fullerenes discussed earlier, which leads to the intermolecular terms of type 1 for $[C_{60}+C_{60}]$ dyad [3] shown in Fig. 1*a*. According to this, the pairwise interaction between fullerene molecules in convenient solutions always leads to the formation of bi-molecular *A+A* or more complex *A+A +A+A* …. [$(C_{60})_n$] homoclusters of fullerenes in the vicinity of the $R^{00}$ minimum on the potential energy curve which we considered earlier. Therefore the PD solutions under ambient conditions should involve conglomerates of clusterized $C_{60}$ molecules as shown schematically in Fig.4, which is experimentally proven in many cases (see for example [20-22]).

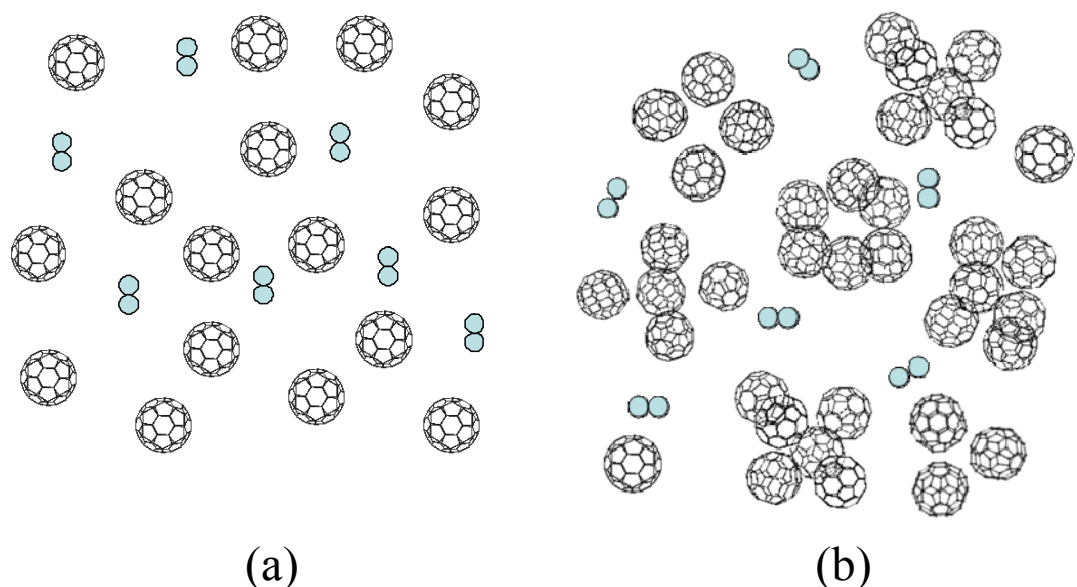

(a)                                              (b)

**Figure 4.** Schematic presentation of an ideal (*a*) and real (*b*) fullerene solution

UBS HF AM1 calculations determine the coupling energy of the pairwise *f-f* interaction for $C_{60}$ as $E_{cpl}^{f-f}$ =-0.52 *kcal/mol*. If remember that $E_{cpl}^{f-o}$=0 in both singlet and triplet state, it becomes clear that oxygen molecules do not interact with either individual fullerene molecule or the molecule clusters so that the total energy of any dyad $[(C_{60})_n\text{-}O_2]$ ( n=1, 2, 3….) is just a sum of those related to the dyad components and is always by 9.93 kcal/mol less in the triplet state due to the difference in the energy of the triplet and singlet oxygen. Therefore the ground state of dyads $^3[(C_{60})_n\text{-}O_2]$ is triplet.

As was shown in the previous section, each pair of fullerene molecules as well as any fullerene cluster of more complex structure formed at the $R^{00}$ minimum is charge transfer complex. The complex absorption band related to $B_2$ phototransitions in Fig.1*a* is located in the UV-visible region. The photoexcitation of either pair or cluster of fullerene molecules within this region produces a pair of molecular ions that quickly relax into the ground state of neutral molecule after the light is switched off. The current calculations have revealed that, oppositely to neutral $C_{60}$, both molecular ions $C_{60}^-$ and $C_{60}^+$ actively interact with oxygen molecule producing coupling energy $E_{cpl}^{-f-o}$ and $E_{cpl}^{+f-o}$ of -10.03 and -10.05 *kcal/mol*, respectively, referring to $^3O_2$ molecule and -0.097 and -0.115 *kcal/mol* in regards to $^1O_2$. Therefore, oxygen molecule is quite strongly held in the vicinity of both molecular ions forming $[C_{60}+O_2]^-$ and $[C_{60}+O_2]^+$ complexes as schematically shown in Fig. 5. UBS HF AM1 calculations for the corresponding dyads show that the complexes are of $^2[C_{60}^-+O_2]$ and $^2[C_{60}^++O_2]$ compositions of the doublet spin multiplicity. Both fullerene ions take the responsibility over the complex spin multiplicity, so that two electrons of the oxygen molecule that were on the service of triplet spin multiplicity of $^3[(C_{60})_n+O_2]$ dyads in the ground state are not more needed for the job and become effectively unpaired thus adding two electrons to the $N_D$ pool of unpaired electrons of complexes $^2[C_{60}^-+O_2]$ and $^2[C_{60}^++O_2]$. The distribution of effectively unpaired electrons of both complexes over their atoms, which displays the distribution of the atomic chemical susceptibility of the complexes, is shown in Fig.6. A dominant contribution of electrons located on oxygen atoms 61 and 62 is clearly seen thus revealing the most active sites of the complexes. It should be noted that these distributions are intimate characteristics of both complexes so that not oxygen itself but both complexes as a whole provide the oxidative effect. The effect is lasted until the complexes exist and is practically immediately terminated when the latter disappear when the light is switched off.

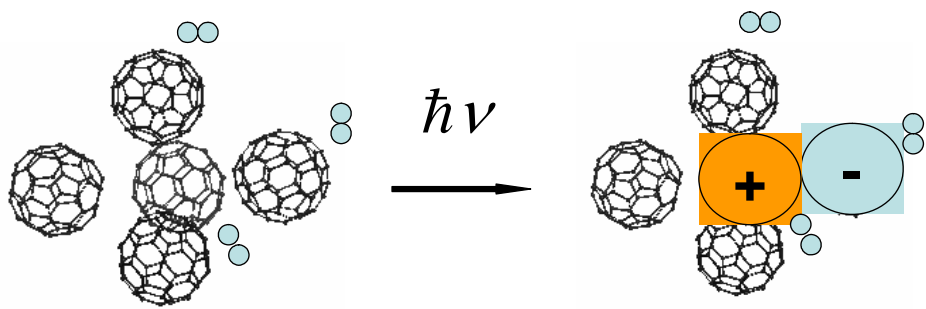

**Figure 5.** The formation of an ionic pair under photoexcitation of $(C_{60})_n$ cluster.

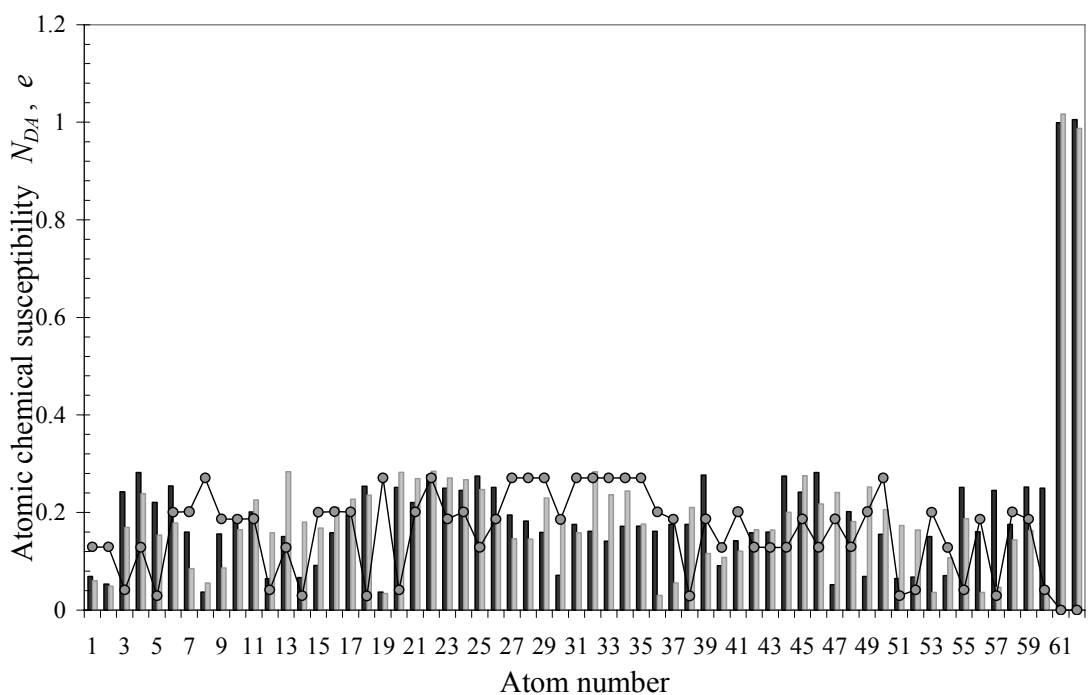

**Figure 6.** Distribution of atomic chemical susceptibility $N_{DA}$ [18] over atoms of $^2[C_{60}^- + O_2]$ (black bars) and $^2[C_{60}^+ + O_2]$ (light gray bars) complexes. Curve with dots plots the distribution over atoms of $^3[C_{60} + O_2]$ complex.

The obtained results make it possible to suggest the following mechanism that lays the foundation of the photodynamic effect of fullerene solutions

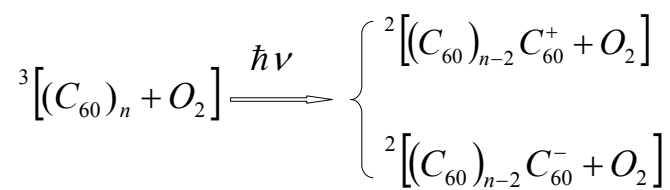

Scheme 2

Here $(C_{60})_{n-2} C_{60}^+$ and $(C_{60})_{n-2} C_{60}^-$ present fullerene clusters incorporating molecular ions that we discussed earlier in regards photostimulated oligomerization. This transformation of the triplet ground state complex into two doublet ones under photoexcitation is accompanied with a spin flip of the oxygen molecule electrons in the presence of fullerene molecule which is shown scematically in Fig. 7. This approach allows attributing phodynamical effect of fullerene solutions to a new type of chemical reactions in the modern spin chemistry.

Since fullerene derivatives preserve D-A properties of the pristine fullerene, Scheme 2 is fully attributable to the latter as well. So that not only $C_{60}$ or $C_{70}$ themselves but their derivatives can be used in PD solutions. Obviously, parameters of the photodynamic therapy should therewith be different depending on the fullerene derivative structure as is actually observed experimentally [19]. Changing solute molecules, it is possible to influence the efficacy of their clusterization, which, in its turn, may either enhance or press the therapeutic effect. The situation appears to be similar to that occurred in the optical behavior of fullerene solution.

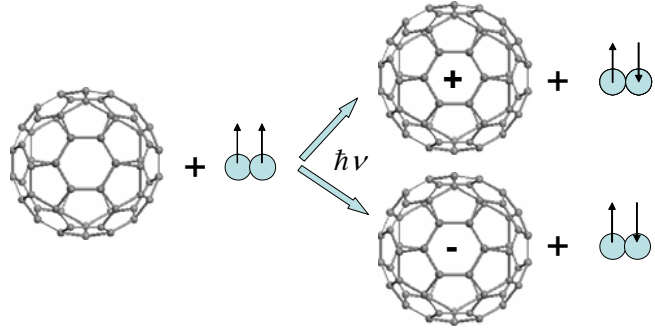

**Figure 7**. Schematic presentation of the spin-flip in oxygen molecule under photoexcitation

## 5. ENHANCEMENT OF OPTICAL RESPONSE IN FULLERENE SOLUTIONS

The third scenario evidently concerns nonlinear optical effects. In terms of nonlinear optics, there are two field effects responsible for enhancement of linear and nonlinear optical properties, namely, field restriction and field resonance. Thus, the arrangement of space in the form of an open (cavity or planar structures) or closed (droplet) nanocell around or near an object to be poled affects a particular spatial distribution of the field around the object, thereby resulting in the amplification of light emitted by this object. On the other hand, when the field of an incident or exiting light wave is in resonance with local excitations of electron–hole plasma of the atomic structure of the cell, additional considerable enhancement of spontaneous and stimulated radiation of the object takes place.

While the former effect has been studied in a wide variety of materials (see review [23] and references therein), the resonance effect was observed largely for nanostructured metals, such as silver and gold, which are characterized by resonance of pumping laser wavelengths with local plasmons in the visible spectral region [24]. However, not only plasmons are capable to generate electron–hole plasma in the medium. Wannier–Mott excitons and/or charge-transfer excitons can do this as well. Excitation of localized excitons, like that of local plasmons, results in significant polarization of the medium and, as such, can cause substantial enhancement of the optical properties of the object placed inside or near a nanocell made from an appropriate material. Nonetheless, investigations in this area are scanty and deal with semiconductor materials described in the Wannier–Mott exciton formalism [25]. In this section, we survey the pioneering results relating to the field resonance effect due to charge-transfer excitons [8, 20, 26].

Field resonance was revealed as enhancement of emission spectra of fullerene solutions in a crystalline toluene matrix at low temperatures. By fullerene is meant $C_{60}$ fullerene and its derivatives. The enhancement of emission spectra is the emergence of unusually intense luminescence in the visible spectral region— blue emission—and the dependence of its intensity (relative to an internal standard) on the wavelength of excitation light $\lambda_{exc}$. It is reasonable to look for explanation of the specifics of the optical behavior of fullerene solutions in terms of the local-field enhancement model [27]. In order for this model to be successfully applied, it was necessary (1) to determine what is the polarizable object in a fullerene solution, (2) to answer the question what is the substance of the solution responsible for excitation of electron–hole plasma, (3) to define the bounds of the plasma resonance frequency range, and (4) to ensure fulfillment of the resonance conditions for local factors. On the basis of the available data, the following answers to these questions were proposed.

(1) The polarizable object in fullerene solutions is nanosized clusters of the solute *sol–sol* composed of fullerene molecules alone and/or *sol–solv* clusters, which include solvent molecules along with fullerenes. Similarly to *sol–sol* clusters, some *sol–solv* clusters, particularly involving toluene, are charge-transfer complexes (see detailed discussion in [20, 26]). The efficiency of clustering depends on the pairwise coupling energy $E_{cpl}\{(F)_2\}$ or $E_{cpl}\{(F)_1(S)_1\}$. Here, $F$ and $S$ are the fullerene and solvent molecules, respectively. The quantity $E_{cpl}$ is a controlling parameter of clustering; thus, the structure of the clusters depends on the chemical structure of the fullerene molecule. Figure 8 exemplifies such clusters simulated by quantum-chemical calculations performed within the AM1 UBS HF approach.

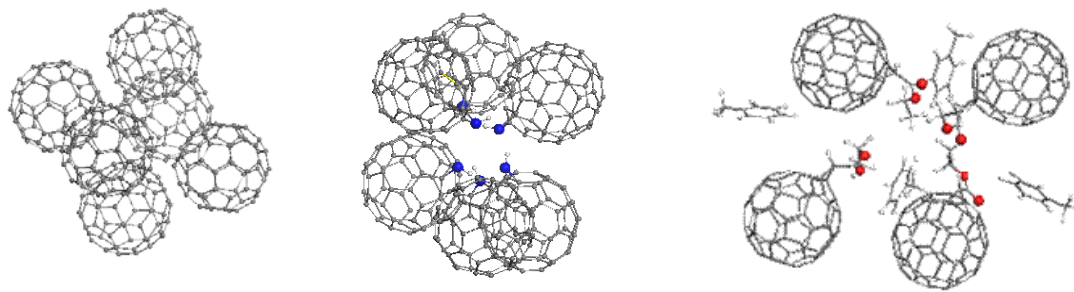

**Figure 8.** Equilibrium structures of clusters (I)$_6$ (*a*), (II)$_6$ (*b*), and (III)$_4$(T)$_4$ (*c*). Symbols I, II, III, and T denote fullerene $C_{60}$, fullerene azyridine [$C_{60}$], ethyl ester of [$C_{60}$]fullerene acetic acid, and toluene molecule, respectively. The relevant $E_{cpl}$ have values of –2.74, –8.42, and –5.81 kcal/mol for (*a*, *b*) *sol–sol* and (*c*) *sol–solv* clusters [28]. Atom marking see in the caption to Fig.3.

(2) The spectrum of excited states of the $C_{60}$ fullerene crystal is the sum of spectra of Frenkel excitons and charge-transfer excitons [29, 30]. The specifics of these two-exciton spectra are their overlapping at the long-wavelength edge corresponding to the absorption spectrum of the crystal in the near IR and visible regions. Accordingly, the energy spectrum of the excited-state clusters of fullerene molecules is the sum of the spectra of localized Frenkel excitons and localized charge-transfer excitons (local charge-transfer states) (Fig. 9). Unlike the spectrum of the crystal, there are two spectra of local charge-transfer states in the solution in accordance with the two types of clusters, *sol–sol* and *sol–solv*. Excitation of these spectra is responsible for the generation of electron–hole plasma, whose polarization provides the amplification of the incident and outcoming electromagnetic wave.

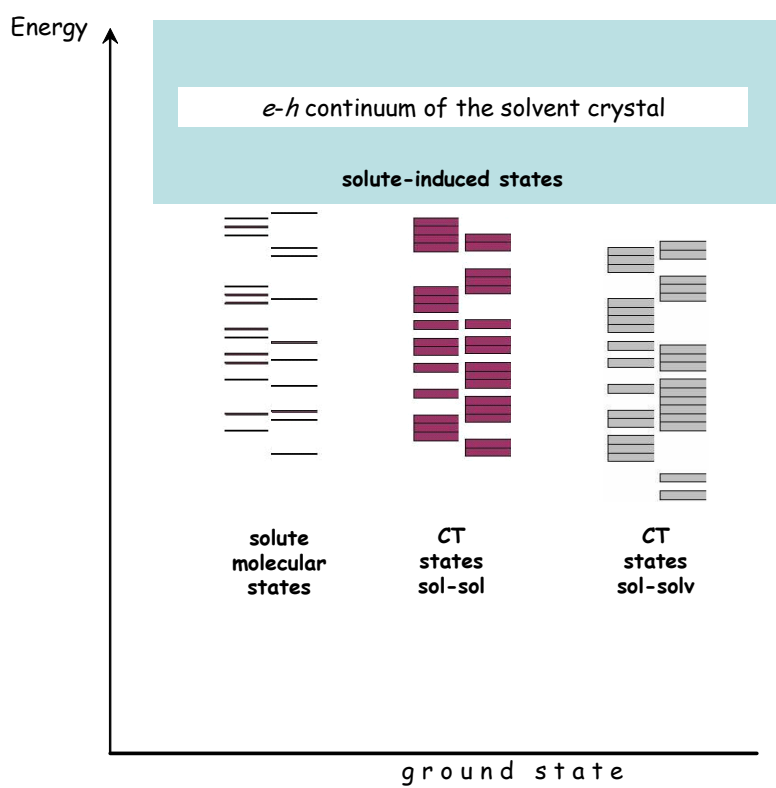

**Figure 9.** Energy diagram for singlet and triplet excited states fullerene in solution [26].

(3) The spectrum of plasma frequencies which meet the resonance conditions of the amplification of incident and exiting light coincides with the spectrum of localized charge-transfer excitons. The location of this spectrum in the visible region causes the emergence of enhanced emission in this exciting light wavelength range. It is this factor that determines the appearance of the intense blue luminescence in addition to the red emission of fullerene crystals and molecular solutions [31, 32]. In all of the examined cases, this emission is either enhanced Raman scattering or enhanced photoluminescence of fullerene clusters and the solvent or a combination of both spectra. An important feature of the blue emission is its dependence upon the structure of the fullerene molecule. Figure 10 presents typical spectra of the blue emission depending on the molecular structure and the excitation wavelength. The presented spectra correspond to the maximum enhancement attained in the experiments [8, 20, 26]. The enhancement is defined as the rise in the relative intensity of the blue emission spectrum with respect to the red emission spectrum of the solution, which plays the role of internal standard and, as such, is averaged in the intensity of the spectra over all solutions in Fig. 10, with a fixed range of variation in $\lambda_{exc}$. The dashed curve at the bottom in Fig. 10*b* shows the spectrum of the density of states of charge-transfer excitons in the $C_{60}$ crystal [33]; from the position of this spectrum, the position of the spectrum of the charge-transfer excited states of *sol–sol* clusters can be inferred. The arrows show the location of radiation $\lambda_{exc}$ of the pumping lasers used in the experiments. As is seen in the figure, the excitation spectrum is situated at the long-wavelength edge of the exciton spectrum.

The blue emission in the spectrum of a solution of fullerene IV (hereinafter solution IV) shown in Fig. 10*a* is very weak and is determined by spontaneous Raman scattering, which does not experience enhancement with a change in $\lambda_{exc}$. This behavior is explained by the nonoccurrence of fullerene clustering in the solution because of the prevalence of repulsion between the molecules [26]. The relative intensity of the Raman spectrum is low in comparison with the red spectrum, it varies insignificantly in different studies and does not exceed 9%. We take this value as the reference level for determination of the enhancement of the blue emission spectrum on passing to other solutions marked by *X*. The gain coefficient *K* is defined below as [34]

$$K = \frac{I_{blueX} / I_{redX}}{I_{blueIV} / I_{redIV}}. \tag{1}$$

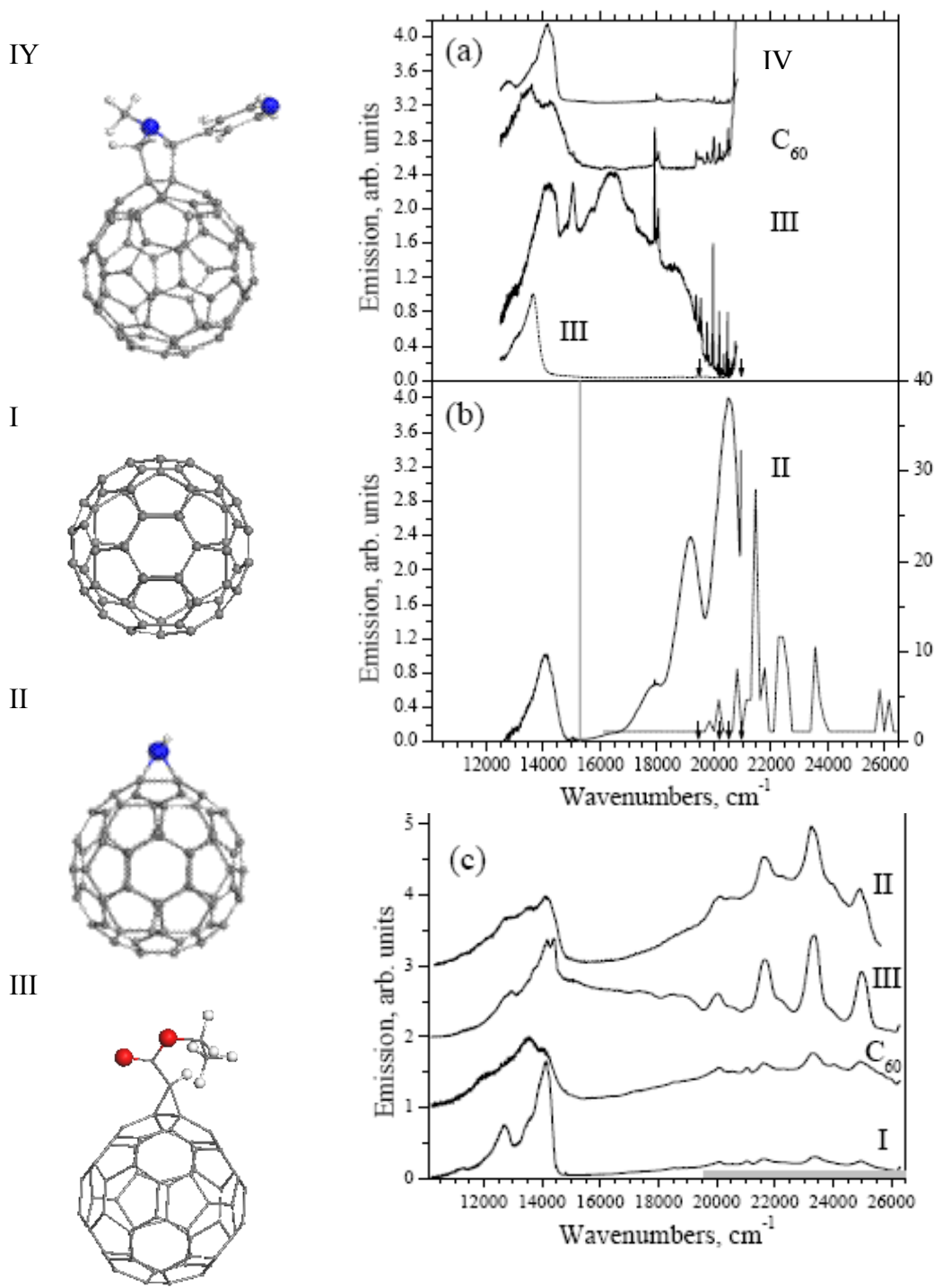

**Figure 10.** (left) Equilibrium structures of fullerene molecules column. IV marks N-methyl-2(4-pyridine)-3,4-[$C_{60}$]fulleropyrrolidine while the rest numbering is the same as in Fig.8. (right) The emission spectra of the solutions in toluene at 80 K excited at $\lambda_{ex}$ of (*a–b*) 476.5 and (*c*) 337.1 nm. Dashed curves show the spectrum of III in carbon tetrachloride (at the bottom in (*a*)) and density of states of charge-transfer excitons of the $C_{60}$ crystal [35]( at the bottom in (*b*). The arrows mark the positions of frequencies of pumping lasers, $\lambda_{exc}$ 514.5, 496.5, 488.0, and 476.5 nm.

Here $I_{blueX}$ and $I_{blueIV}$ are the overall intensities of the blue spectra of solutions X and IV, respectively, and $I_{redX}$ and $I_{redIV}$ refer to the red spectra of these solutions.

The blue emission of solution I in Fig. 10*a* is the solvent-enhanced Raman scattering (SOERS) [26]. The amplification of the relative intensity is by factor 2.5 with a change in $\lambda_{exc}$ from 514.5 to 476.5 nm. The SOERS spectrum itself makes up 24% of the red spectrum at $\lambda_{exc}$ = 476.5 nm. Thus, the total gain coefficient *K* is 2.7.

The blue emission of solution II in Fig. 10*b* is fullerene-enhanced Raman scattering (FERS) due to light scattering by *sol–sol* clusters of fullerene II [8]. Upon passing from 514.5 to 476.5 nm, the relative intensity increases by a factor of ~4. At $\lambda_{exc}$ = 476.5 nm, the $I_{blueIII} / I_{redIII}$ ratio is 66; thus, the gain coefficient is *K*=733. In the case of solutions I and II, the maximal enhancement was attained under conditions when the wavelength of excitation light corresponded to the maximal penetration into the spectrum of *sol–sol* clusters and the blue emission spectra overlapped it.

For solution III, the blue emission spectrum shown in Fig. 10*a* is composed of the SOERS of toluene (narrow lines) against the background of enhanced broad-band photoluminescence of *sol–sol* fullerene clusters (fullerene-enhanced luminescence, FEL) [26]. The overall emission intensity increases by a factor of 3 as $\lambda_{exc}$ is changed from 514.5 to 476.5 nm. At $\lambda_{exc}$ = 476.5 nm, $I_{blueIVI} / I_{redIV}$ is 3.25 and the gain coefficient is *K*= 36. For this solution, as in the previous cases, the maximum enhancement corresponds to $\lambda_{exc}$ = 476.5 nm; i.e., the excitation falls into the energy spectrum of *sol–sol* clusters; however, the substantial portion of the FEL spectrum occurs at longer wavelengths in comparison to this spectrum. It might seem that one of the fundamental conclusions of the local field theory, which suggests that the emission spectrum in the case of photoluminescence necessarily overlaps the spectral region of excitation of plasma frequencies [35], is violated in this case. However, the specific role played by the solvent is manifested in this example. The emergence of FEL is due to the fact that both *sol–sol* and *sol–solv* clusters can be formed in solution III (see Fig. 8*c*). The energy spectrum of the latter clusters is displaced to longer wavelengths by ~0.5 eV relative to the spectrum of *sol–sol* clusters and, occurring under the FEL spectrum and thus compensating for its Stokes shift, provides necessary conditions for the enhancement of photoluminescence of *sol–sol* clusters [20, 26]. In contrast, the spectrum of *sol–solv* clusters in solution II remains almost in the same position relative to the spectrum of *sol–sol* clusters; as a result, there is no 'resonance support' for their photoluminescence so that the enhancement conditions are met only for light scattering, a factor that is responsible for the appearance of FERS.

Another type of blue emission is illustrated in Fig. 10*c*. This is enhanced luminescence of toluene (solvent-enhanced luminescence, SEL) [8, 20, 26] at $\lambda_{exc}$ = 337.1 nm. As is seen from the figure, the SEL spectrum falls into the region of energy spectrum of *sol–sol* clusters, thereby resulting in its enhancement. The relative intensity of this spectrum increases on passing from solution IV to solutions III and II. This dependence reflects the increase in the efficiency of clustering in the solutions, which is determined by a rise in $E_{cpl}$. If the spectrum of solution IV is taken to be zero reference level, as in the previous case, the gain coefficient according to Eq.(1) for solution II will be 6. As follows from the foregoing, the appearance of intense blue emission of a fullerene solution is undoubted evidence for the clustering of fullerene molecules. The variety of the blue emission components highlights a strong competition between the *sol–sol* and *sol–solv* clusters that influences the spectrum formation. Thus, a small gap between the energies of the *sol–sol* and *sol–solv* clusters favors the manifestation of FERS, whereas high values of the energy gap will be favorable for FEL.

(4) The arguments presented in clauses (1)–(3) in favor of the electromagnetic nature of enhancement of light wave fields refer to necessary conditions for observation of enhanced spectral effects. However, they are insufficient [27] without simultaneous fulfillment of the resonance conditions for the local factors that concern the cluster shape and follow from the relationship

$$\text{Re}[\varepsilon(\omega_R)Q_1(\xi_0) - \xi_0 Q_1'(\xi_0)] = 0. \tag{2}$$

Here, $\xi_0$ is the parameter of the geometric shape of the object and $Q_1$ and $Q_1'$ are the second-order Legendre polynomial and its derivative, respectively. In the case of spherical particle, this expression is reduced to the form

$$\text{Re}\,\varepsilon(\omega_R) + 2 = 0, \tag{3}$$

thereby making it possible to estimate the order of magnitude of the real dielectric function, which determines the value of the resonance frequency $\omega_R$. For most molecular substances, $\text{Re}\,\varepsilon(\omega)$ is always positive and its value is high. In contrast, $\text{Re}\,\varepsilon(\omega)$ in $C_{60}$ crystal can vary from +3 to –5 in the $\lambda_{exc}$ interval of interest, 2.4–2.6 eV, as is seen from the data in Fig. 11. It seems to be evident that this very fact, exceptional for molecular compounds, provides the fulfillment of resonance or at least close-to-resonance conditions for not only the $C_{60}$ crystal, but also solutions of $C_{60}$ and its derivatives.

The results discussed above have shown that nanosize fullerene clusters in solutions can act as light wave enhancers, like gold or silver colloid particles. The question arises of what is the attainable magnitude of enhancement of *K* in this case. It is known that *K* for silver colloid particles is $10^4$–$10^6$. In terms of the local field theory, enhancement of a light wave is due not only to the fulfillment of the resonance conditions (2) and (3) for $\text{Re}\,\varepsilon(\omega)$, but also to other factors. Thus, at the resonance frequency, the intensity of the spectrum is controlled by the dielectric loss factor of the object, causing its dependence [35]

$$I \approx (\mathrm{Im}\,\varepsilon(\omega_R))^{-4}.\qquad(4)$$

The difference determined by these two circumstances for fullerene clusters and silver particles can be revealed by comparative analysis of the dielectric losses of the $C_{60}$ fullerene crystal and silver shown in Fig. 11. As seen in the figure, $\mathrm{Re}\,\varepsilon(\omega)$ values coincide for the both enhancers in the region of our interest, where the values of $\mathrm{Im}\,\varepsilon(\omega)$ differ by a factor of about 15 in favor of silver particles; that is, there is a $\sim 10^4$-fold difference in amplification. According to these tentative data, the maximal enhancement attainable with the use of fullerene clusters cannot be higher than $10^2$. Of course, these estimates are quite approximate, but they show the order of magnitude. In the experiments describe above, the coefficient $K$ varied from 2.6 in solution I to 733 in solution II. The latter value appears to be close to the limit. A high value of $K$, along with other characteristics, indicates the undoubted readiness of this solution for effective NLO applications.

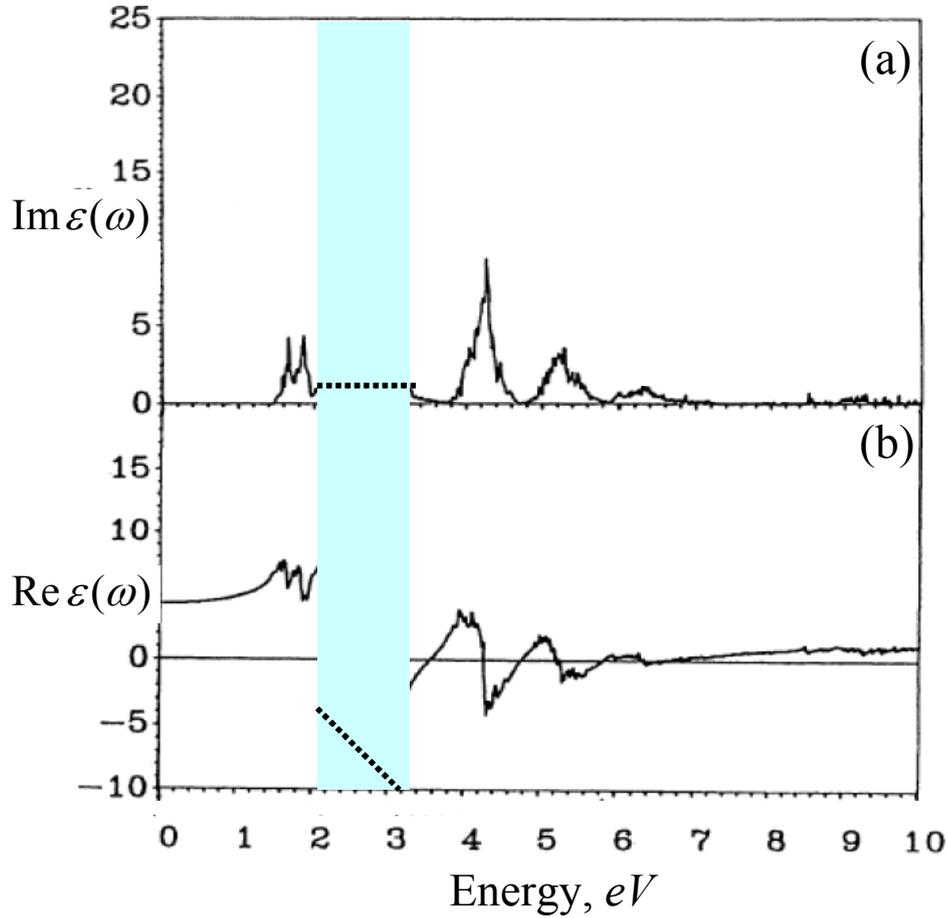

**Figure 11.** Dielectric loss spectrum of the $C_{60}$ crystal: $\mathrm{Im}\,\varepsilon(\omega)$ (*a*) and $\mathrm{Re}\,\varepsilon(\omega)$ (*b*) (calculations) [36]. The shaded area corresponds to the visible region of the spectrum. The dashed segments show fragments of the dielectric loss spectrum of silver [37].

As applied to the NLO properties, the investigations [20, 26, 34] have shown that the position of both the energy spectrum of charge-transfer excitons localized on fullerene clusters and the energy range of resonance conditions of resonance conditions of local factors in the visible spectral region makes fullerene-doped matrices excellent objects providing for the enhancement of NLO properties in this region. The discovery of intense blue emission can be convincing evidence that this medium is suitable for NLO applications. Thus, the blue spectrum becomes an empirical test whose application makes it possible to select an NLO medium with required properties. An auxiliary test can be poor solubility of fullerene in the matrix of interest, a property that generally favors the clustering of the solute [38]. Along with these two empirical tests, there is a computational test based on the determination of the *sol–sol* and *sol–solv* pairwise coupling energy $E_{cpl}$. If this energy exceeds 0.5 kcal/mol by absolute value, the solution can be considered a candidate for an effective NLO medium. The validity of these criteria for selection of an active NLO medium has been verified recently with a $C_{70}$ fullerene solution in cyanobiphenyl [39] as an example representing an NLO medium widely used for effective incident-light intensity limiters, displays, and bidirectional diffraction elements [40].

## 6. CONCLUSION

Describing the above three photosensitive effects, three different languages were used reflecting different branches of science equipped by different sets of terms and concepts. And nevertheless, it has been quite evident that the effects could not exist if there was no interaction between fullerene molecules. All these effects are not monomolecular and involve a group of, at least of two molecules and are a consequence of a peculiar intermolecular interaction occurred. As shown, the formation of a positive-negative ionic pair of fullerene under the photoexcitation is a key point of the interaction. The community of this process for all observed effects makes it possible to join them under a common umbrella of nanophotonics of fullerene that is a peculiar intermolecular phenomenon [41].